# 729 new measures of economic complexity

## (Addendum to Improving the Economic Complexity Index)

Saleh Albeaik[1], Mary Kaltenberg[2,3], Mansour Alsaleh[1], César A. Hidalgo[2]

[1] Center for Complex Engineering Systems, King Abdulaziz City for Science and Technology
[2] Collective Learning Group, The MIT Media Lab, Massachusetts Institute of Technology
[3] UNU-MERIT, Maastricht University

**Abstract:**
Recently we uploaded to the arxiv a paper entitled: Improving the Economic Complexity Index. There, we compared three metrics of the knowledge intensity of an economy, the original metric we published in 2009 (the Economic Complexity Index or ECI), a variation of the metric proposed in 2012, and a variation we called ECI+. It was brought to our attention that the definition of ECI+ was equivalent to the variation of the metric proposed in 2012. We have verified this claim, and found that while the equations are not exactly the same, they are similar enough to be our own oversight. More importantly, we now ask: how many variations of the original ECI work? In this paper we provide a simple unifying framework to explore multiple variations of ECI, including both the original 2009 ECI and the 2012 variation. We found that a large fraction of variations have a similar predictive power, indicating that the chance of finding a variation of ECI that works, after the seminal 2009 measure, are surprisingly high. In fact, more than 28 percent of these variations have a predictive power that is within 90 percent of the maximum for any variation. These findings show that, once the idea of measuring economic complexity was out, creating a variation with a similar predictive power (like the ones proposed in 2012) was trivial (a 1 in 3 shot). More importantly, the result show that using exports data to measure the knowledge intensity of an economy is a robust phenomenon that works for multiple functional forms. Moreover, the fact that multiple variations of the 2009 ECI perform close to the maximum, tells us that no variation of ECI will have a performance that is substantially better. This suggests that research efforts should focus on uncovering the mechanisms that contribute to the diffusion and accumulation of productive knowledge instead of on exploring small variations to existing measures.

# A Tale of Two Measures

In 2009 we published a paper[1] in PNAS that proposed a metric to estimate the knowledge intensity of countries and products by looking at the structure of the network connecting countries to the products they export. The formula assumed that the knowledge intensity of a country (its economic complexity) was equal to the average knowledge intensity of the products it exported. Conversely, the knowledge intensity of a product was equal to the average knowledge intensity of the countries exporting it.

Mathematically, this intuition can be formalized by having data on which country exports each product, and simply setting the knowledge intensity of a country (Kc) to be equal to the average knowledge intensity of a its products (Kp), and the knowledge intensity of a product (Kp) to be equal to the average knowledge intensity of the countries exporting it (Kp). If Mcp is a matrix telling you which countries export which product[i], then:

$$K_c = \frac{\sum_p M_{cp} K_p}{\sum_p M_{cp}}$$

$$K_p = \frac{\sum_c M_{cp} K_c}{\sum_c M_{cp}}$$

This circular equation can be solved by taking values of Kc and Kp and feeding them to each other iteratively. This can be done by setting an iteration between Kc(n+1) and Kp(n) and Kp(n+1) +Kc(n).

---

[i] You define the countries that export a product, as the countries that export more than what you expect based on their size and the size of the market of a product. If Xcp are the exports of country c in product p, Xc are the total exports of a country, Xp are the total exports on a product, and X are the total exports of the world, then Mcp =1 if Xcp > XpXc/X

$$K_c(n + 1) = \frac{\sum_p M_{cp} K_p(n)}{\sum_p M_{cp}}$$

$$K_p(n + 1) = \frac{\sum_c M_{cp} K_c(n)}{\sum_c M_{cp}}$$

We also published this measure of knowledge intensity (in 2011) in a book[2] (first on the web, and then on MIT press) which combined the use of knowledge intensity to predict growth with the study of knowledge diffusion among related products (the idea of the product space)[3].

In 2012 a group published a variation of our 2009 formula in a paper entitled: "*A New Metrics for Countries' Fitness and Products' Complexity*". Their variation, which they did not interpret in terms of knowledge intensity, replaced the first average for a sum, and the second average for the inverse of the sum of the reciprocals, creating a similar formula:

$$K_c = \sum_p M_{cp} K_p$$

$$K_p = \frac{1}{\sum_c M_{cp} \frac{1}{K_c}}$$

In that paper, the team also tried an alternative form (which they define mathematically in an endnote). In that second alternative, which they called *extensive fitness*, Mcp was not a discrete matrix connecting countries to their more relevant exports, but a matrix with the share that each product represents in a country's total exports. If the exports of country c in product p are Xcp, then, they replaced Mcp for:

$$M_{cp} = \frac{X_{cp}}{\sum_p Xcp}$$

Yet, to obtain the formula of our working paper Mcp needs not to be replaced by $\frac{X_{cp}}{\sum_p X_{cp}}$, but simply by Xcp (the exports of a country in a product). If we replace Mcp by Xcp in the equations proposed in[4]:

$$K_c = \sum_p X_{cp} K_p$$

$$K_p = \frac{1}{\sum_c \frac{X_{cp}}{K_c}}$$

And include the second equation (Kp) into the first one (for Kc) we obtain:

$$K_c = \sum_p \frac{X_{cp}}{\sum_c \frac{X_{cp}}{K_c}}$$

This derived equation is equivalent to the first term of an equation we had in a recent working paper comparing and exploring a variation in a metric of knowledge intensity[5]. Yet, we did not realize that the equation below, was the equivalent of having introduced Kp into the equation of Kc and used Xcp instead of Mcp in[4], because we did not go through that derivation to arrive at the short formula. Instead, we came through the route of knowledge intensity and considered that the more knowledge intense products are those in which it is harder to generate each dollar of exports. So we needed to correct a country's total exports ($X_c = \sum_p X_{cp}$), by how knowledge intense was the export of each product. And since it is harder to enter the market of knowledge intense products, few countries would have a large market share on the knowledge intense products. So we can take the average market share of a country in a product $\sum_c \frac{X_{cp}}{K_c}$ as a measure that is the inverse of its knowledge intensity. That gives us:

$$X_c = \sum_p \frac{X_{cp}}{\sum_c \frac{X_{cp}}{X_c}}$$

This equation for Xc is equivalent to the equation obtained after combining the equations for Kc and Kp. in the second method introduced in[4].

So it was our oversight not to have seen the functional equivalence between the equation for Xc and the equation you get by including Kp into Kc in the second method presented in[4] and replacing Mcp by Xcp.

We acknowledge this oversight and are adding this addendum to the working paper. Yet, more importantly, this motivated us to explore a more interesting question. That is: how easy it is to create a variation of our 2009 metric of economic complexity that works?

**729 measures of economic complexity**

How many variations of ECI produce a measure of knowledge intensity, or economic complexity, that is predictive of future economic growth? To explore this, consider the following unifying framework, which contains our 2009 measure[1] (the economic complexity index) and also, the 2012 variations proposed in[4]:

$$K_c = \frac{\left(\sum_p M_{cp} K_p^{\alpha}\right)^{\gamma}}{\left(\sum_p M_{cp}\right)^{\varepsilon}}$$

$$K_p = \frac{\left(\sum_c M_{cp} K_c^{\beta}\right)^{\delta}}{\left(\sum_c M_{cp}\right)^{\theta}}$$

In this formula, the 2009 economic complexity index (ECI) is obtained when all exponents are equal to 1 ($\alpha = \beta = \gamma = \delta = \varepsilon = \theta = 1$). To obtain the 2012 fitness formula we need to set $\alpha = \gamma = 1; \beta = \delta = -1; \varepsilon = \theta = 0$. But what about other combinations? For instance, when all coefficients would be equal to -1? Would these combinations also generate measures of knowledge intensity that are predictive of future economic growth?

We consider three possible values [1, 0, -1] for the coefficients ($\alpha$ to $\theta$) (we could consider more, but it is beyond the point). With these three values we obtain a set of 729 possible variations of ECI ($3^6$=729). We note that some of this variations are equivalent, for instance, when $\alpha = 0$, the case when $\varepsilon = 1$ and $\gamma = 1$ is equivalent to the case when $\varepsilon = -1$ and $\gamma = -1$. Yet, these anecdotal symmetries should not affect the general point we will demonstrate. Also, we could explore more combinations if we consider replacing Mcp with Xcp/Xc, or Mcp=Xcp, etc. Yet, exploring these extra variations would not add much conceptually if within the first set of variations we find many with a similar predictive power than ECI and the variations proposed in 2012.

So we construct the 729 combinations of the formula considering Mcp and run a 10-year growth regression for each of them to identify the sets of parameters that are predictive of future economic growth. In all cases, we normalize the variables by subtracting their respective means and dividing by their standard deviations. We then use the following baseline growth model to test the predictive power of each variation of ECI:

$$G(t, t + 10) = ECI(t) + GDPpc(t) + POP(t) + C$$

Where *G(t,t+10)* is the compound annualized growth rate in ten years, ECI is one of the 729 variations of ECI, GDPpc is the log of the GDP per capita of a country, and POP(t) is the log of the population of the country.

Figures 1 and 2 show a couple of examples of the $R^2$ obtained for this baseline regression for each of the 729 variations of ECI as a function of an index running from 1 to 729. Since we set up the iteration to loop each variable from -1 to 0 to 1, the Original ECI is the last variation (1,1,1,1,1,1) and the 2012 variation proposed in[4] is variation 545. Figure 1 shows the $R^2$ associated with each variation in a growth regression predicting annualized growth between 1995 and 2005, using 1995 data. Figure 2 does the same for data between 1998 and 2008. In both cases we observe that a large number of variations have a predictive power that is similar to that of the original ECI and of the variation proposed in 2012. In fact, in the second case (Figure 2), almost all regressions have a predictive power between 16% and 18%, indicating that most variations work for that pair of years[ii].

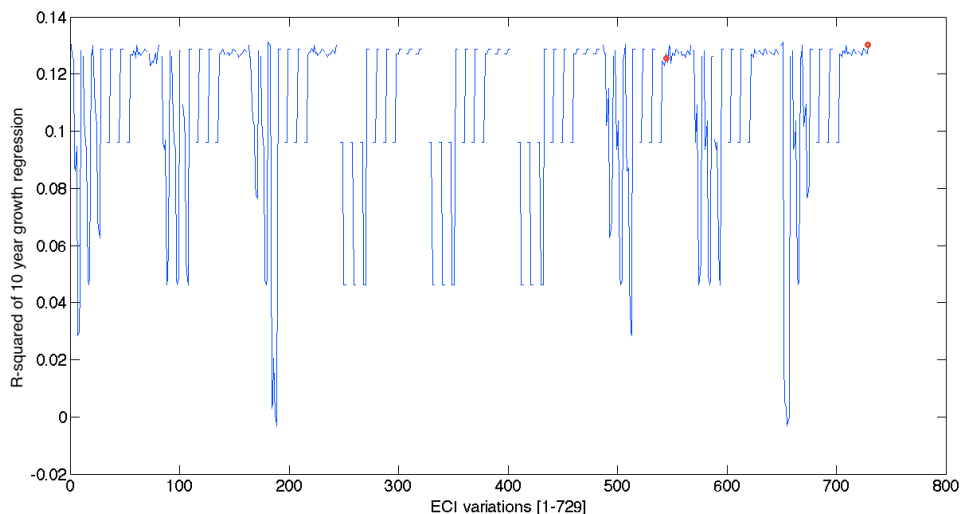

**Figure 1** $R^2$ coefficients of growth regressions for annualized growth rate between 1995 and 2005 considering each of the 729 possible variations in ECI. Original ECI is the last combination (1,1,1,1,1,1) (indicated with a red circle), and the 2012 variation (1,-1,1,-1,0,0) proposed in[4] is variation number 545 (also indicated with a red circle).

---

[ii] This means that for that year the contribution of ECI is small and most the R2 is attributed to the Solow term of the regression (the income term).

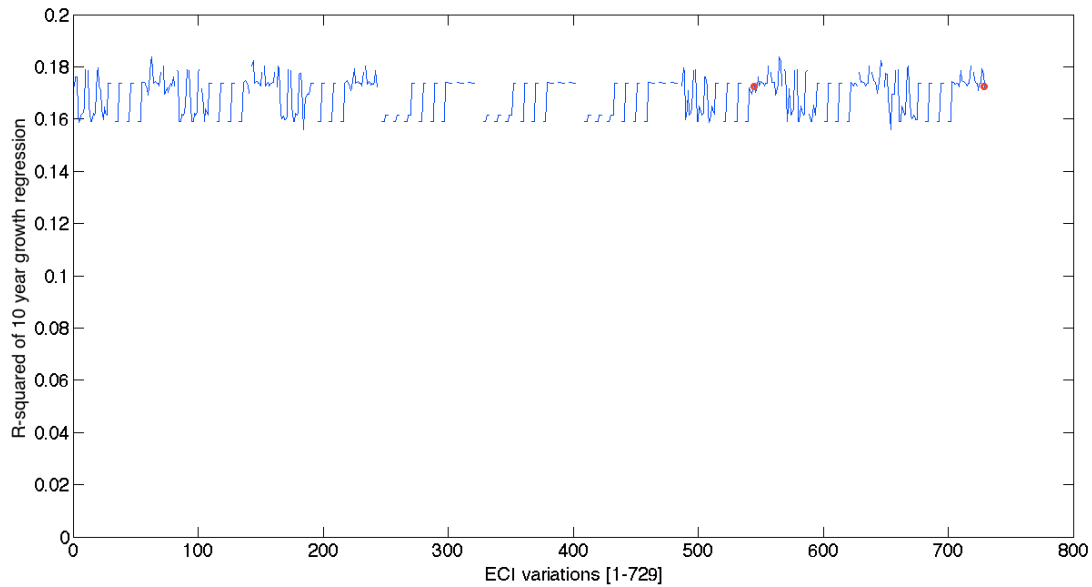

**Figure 2** $R^2$ coefficients of growth regressions for annualized growth rate between 1998 and 2008 considering each of the 729 possible variations in ECI. Original ECI is the last combination (1,1,1,1,1,1) (indicated with a red circle), and the 2012 variation (1,-1,1,-1,0,0) proposed in[4] is variation number 545 (also indicated with a red circle).

But how many of these variations work? We run 10-year annualized growth regressions for all years for which we have data starting from 1988 and count the number of times a variation provides an accuracy that is within 90% or 80% of the maximum. Note that neither ECI nor the 2012 variation are necessarily the maximum. We find that 29% of variations have a predictive power that is within 90% of the maximum, and that more than 33% of variations have a predictive power that is within 80% of the maximum. This shows two things. First, it shows that once the idea of using trade data and iterative averages to measures economic complexity was out, coming out with a variation like the one introduce later in[4] was trivial, since flipping some coefficients randomly gives roughly a 1 out of 3 chance of getting a comparatively good measure in terms of its ability to predict future economic growth. The second, and more important result, is that measuring economic complexity using exports data and iterative averages appears to be a much more robust phenomenon than originally thought, since a wide array of measures captures information similar to the one obtained by the economic complexity index. Finally, the fact that there are many solutions near the maximum, tells us that no variation of a measure of economic complexity—obtained through this methods—will perform substantially better than others, since there are hundreds of variations with an almost identical performance.

Figure 3 shows some example of top and bottom country lists generated with until now unpublished variations of the economic complexity index:

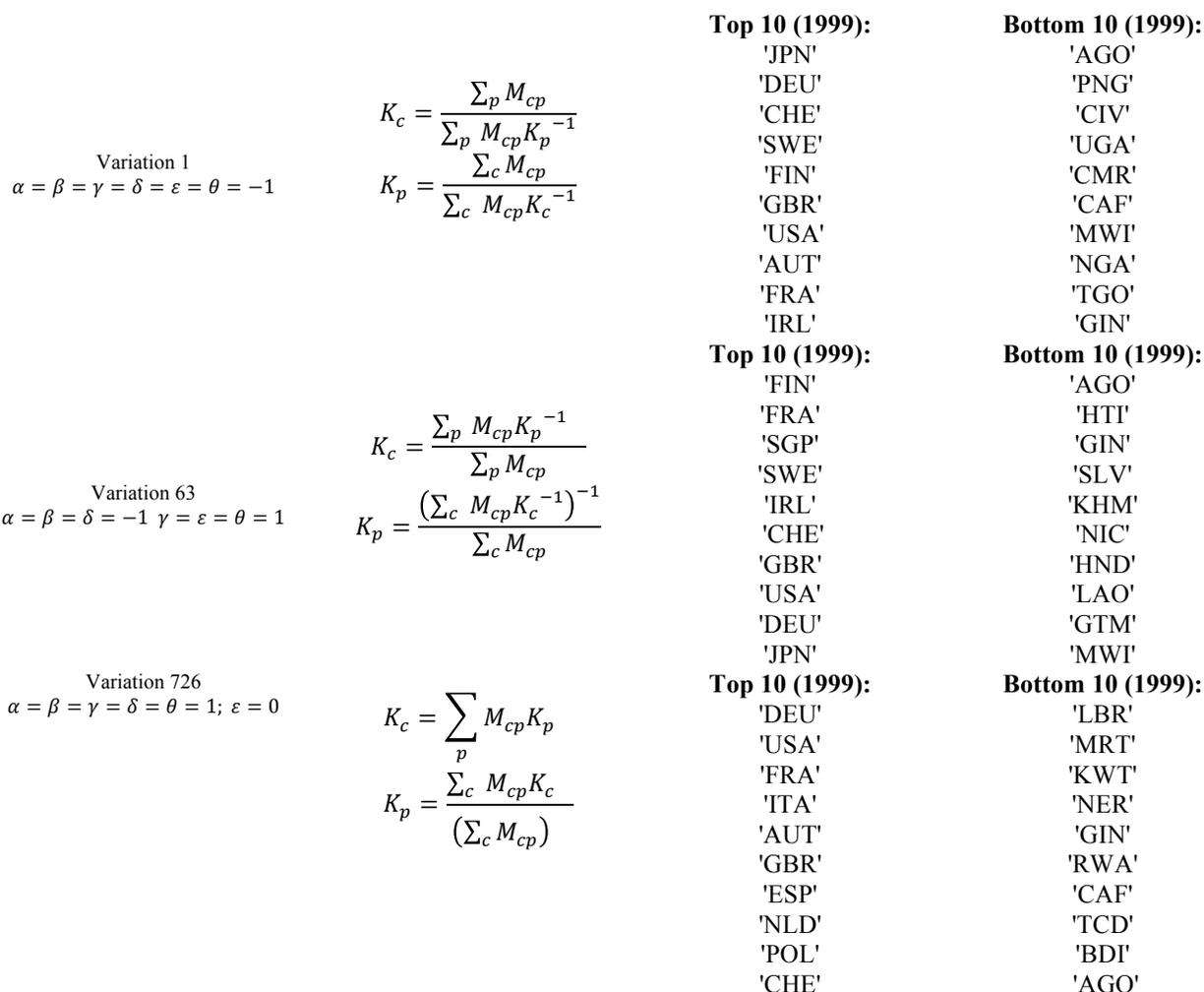

**Figure 3.** Three example variations of the economic complexity index that perform well at predicting future economic growth and that produce sensible rankings.

But does this mean that all variations work? We then explore the variations that work consistently by looking at ten-year growth regressions considering all starting years from 1985 to 2000. Figure 4 looks at the number of times each of these variations produced a measure of economic complexity that is a positive and significant predictor of future economic growth (with a p-value for the regression coefficient of p<0.01). We can clearly see that some variations work better than others (although there are many variations that work).

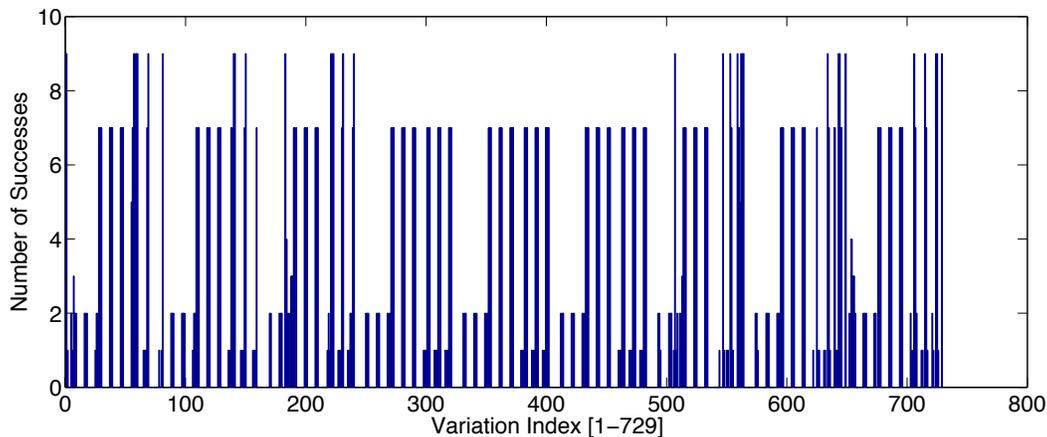

**Figure 4.** Number of times each variation of ECI provided statistically significant coefficients in 15 growth regressions.

The best performing variations work in 9 out of 15 cases, but many other variations provide a significant coefficient in as many as 7 regressions. For comparison, replacing these measures of economic complexity with random numbers provides 0 successes. This tells us that measuring economic complexity works, but that there are a wide number of variations that behave as suitable measures of economic complexity.

Since, we are interested in variations that predict economic growth, we choose variations that are predictive of economic growth with a significance of p<0.01 in most regressions (9 out of 15). We find 32 variations that satisfy this criterion, including the original 2009 ECI, but not the 2012 variation proposed in[4].

Figures 5 **a** and **b** compare the coefficients ($\alpha\ to\ \theta$) of all variations with that of the 32 variations that always produced positive and significant results. In these figures, each row represents a sequence from $\alpha\ to\ \theta$,

blue indicates a -1, green a zero, and red indicates a +1. Figure 5 a shows the 729 variations explored. Figure 4 b shows the 59 variations that had coefficients that were significant predictors of economic growth in all 15 regressions.

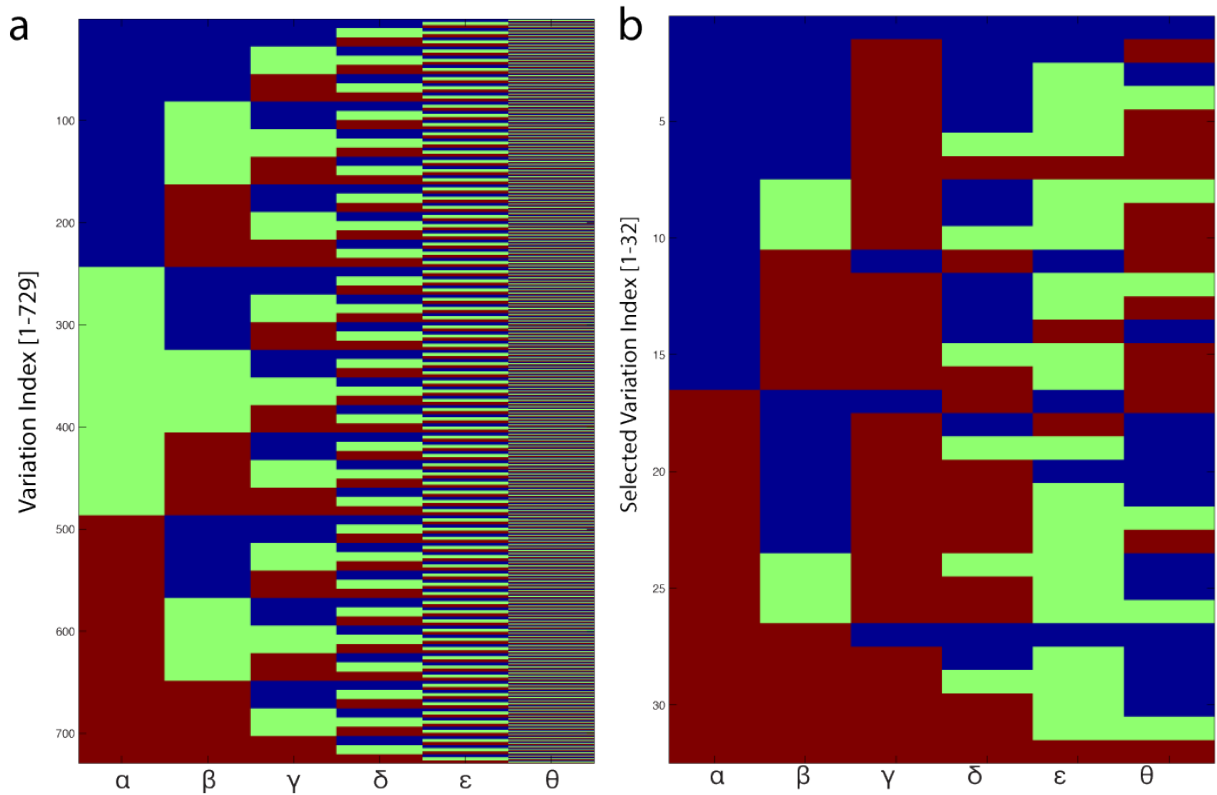

**Figure 5**. **a** Summary of all variations explore. Red indicates a coefficient of 1, green of zero, and blue, of -1. **b** Coefficients of the 32 variations that produced statistically significant coefficients in 9 out of the 15 growth regressions.

From figure 5 **b** we can clearly read a few characteristics of the variations that are most predictive. First, variations with $\alpha = 0$ and $\gamma = 0$ are excluded, meaning that a necessary condition to have a working measure of economic complexity is to couple the equation for countries with that of products. This reinforces the idea that measures of diversity that fail to take into account the sophistication of products are not good measures of the knowledge intensity of an economy. Second $\beta$ is rarely zero, meaning that the Kc term should survive when we include the equation of Kp into that of Kc. Third, when $\alpha = 1$ we are more likely to observe $\theta = -1$ and $\delta = 1$; and when $\alpha = -1$ we are more likely to observe $\theta = 1$ and $\delta = -1$. Also $\gamma$ is almost always 1 and $\varepsilon$ is mostly equal to 0.

These results suggest two alternative forms that reduce to the same equation once we include the equation of Kp into that of Kc. The two forms are:

$$K_c = \sum_{p'} M_{cp'} K_{p'}$$

$$K_p = \sum_{c} M_{cp} K_c^{\beta} \sum_{c'} M_{c'p}$$

and

$$K_c = \sum_{p'} M_{cp'} (1/K_{p'})$$

$$K_p = \frac{1}{\sum_{c} M_{cp} K_c^{\beta} \sum_{c'} M_{c'p}}$$

Which after inserting the equation of Kp into Kc gives respectively the same equation for Kc:

$$K_c = \sum_{p'} M_{cp'} \sum_{c'} M_{c'p} \sum_{c} M_{cp} K_c^{\beta}$$

Of course, since $\beta$ is rarely zero this reduces to two alternatives $\beta = 1$ and $\beta = -1$ which are represented by variations 60, 222, 562, and 724.

But do these two variations provide a significant boost in predictive performance over ECI? The answer is no. When we compare the $R^2$ of the regressions involving variations 60, 22, 562, and 724, with that of ECI (variation 729), we find that the difference in $R^2$ between the predictions obtained by these four variations (technically two, since they only differ on $\beta = 1$ and $\beta = -1$) is always less than 1% and they are statistically insignificant.

But how can so many variations work? The answer is simple. Most variations are highly correlated. In fact, 40% of variations have a correlation with the 2009 ECI that is as high as $R^2$=50% or higher.

Together, these results tell us that attempts[4] to improve on the 2009 versions of the economic complexity index[1], including our recent attempt[6], are futile and completely miss the point. Attempting to improve on the measure before exploring the landscape was our biggest oversight. We now know that many variations work well, and that the seminal idea holds (to create measures of the knowledge intensity of economies one needs to create measures of economic diversity corrected by the knowledge intensity of products). Changing a few coefficients in search for a better measure doesn't amount to even a marginal contribution, since it is after all, a misguided dead-end question. Finding the original 2009 ECI was like finding an island full of coconuts. Finding a variation after that is like finding a coconut in that island.

More importantly, these results also tell us about more fruitful directions for this literature. Now we know that many measures perform similarly, and this tells us that measuring the knowledge intensity of economies is a robust phenomenon. So what we should focus instead? Instead, we should continue to focus both, on the mechanisms that facilitate knowledge diffusion and accumulation, and on additional implications of the knowledge intensity of economies, like its connection with institutions and income inequality[7,8], and the mechanisms that mediate knowledge transfer among related products[2], industries[9,10,11,12], neighboring countries[13], regions[12,14], research areas[15], and technologies[16,17].

## Acknowledgements:

We acknowledge the support of Tarik Roukny and Bogang Jun, who commented on multiple iterations of this draft.